\documentclass[twocolumn]{aastex61}
\usepackage{mhchem}
\usepackage{multirow}
\usepackage{color}

\newcounter{defcounter}
\newenvironment{reaction}{%
\addtocounter{equation}{-1}
\refstepcounter{defcounter}
\renewcommand\theequation{R\,(\thedefcounter)}
\begin{equation}}
{\end{equation}}

\shorttitle{Rate Constants for H$_{2}$CO + H}
\shortauthors{Song et al.}
\begin{document}

\title{Tunneling rate constants for H$_{2}$CO + H on amorphous solid water surfaces}
\correspondingauthor{Johannes K\"astner}
\email{kaestner@theochem.uni-stuttgart.de}

\author{Lei Song}
\author{Johannes K\"astner}
\affiliation{Institute for Theoretical Chemistry, University of Stuttgart, \\
             Pfaffenwaldring 55, 70569 Stuttgart, Germany}

\begin{abstract}
Formaldehyde (H$_{2}$CO) is one of the most abundant molecules observed in the
icy mantle covering interstellar grains. Studying its evolution can contribute
to our understanding of the formation of complex organic molecules in various
interstellar environments.  In this work, we investigated the hydrogenation
reactions of H$_{2}$CO yielding CH$_{3}$O, CH$_{2}$OH and the hydrogen
abstraction resulting in H$_{2}$ + HCO on an amorphous solid water (ASW)
surface using a quantum mechanics/molecular mechanics (QM/MM) model. The
binding energies of H$_{2}$CO on the ASW surface vary broadly from 1000~K to
9370~K. No correlation was found between binding energies and activation
energies of hydrogenation reactions.  Combining instanton theory with QM/MM
modeling, we calculated rate constants for the Langmuir--Hinshelwood and the
Eley--Rideal mechanisms for the three product channels of H + H$_{2}$CO
surface reactions down to 59~K.  We found that the channel producing
CH$_{2}$OH can be ignored owing to its high activation barrier leading to
significantly lower rates than the other two channels.  The ASW surface
influences the reactivity in favor of formation of CH$_{3}$O (branching ratio
$\sim$80\%) and hinders the H$_{2}$CO dissociation into H$_{2}$+HCO.  In
addition, kinetic isotope effects are strong in all reaction channels and
vary strongly between the channels.  Finally, we provide fits of the rate
constants to be used in astrochemical models.
\end{abstract}


\section{Introduction} \label{sec:intro}




Formaldehyde (H$_{2}$CO) and complex organic molecules (COMs) are broadly
observed in cold dense molecular clouds and high- or low-mass protostars. As a
precursor to sugars formed through the formose reaction or as a precursor to
amino acids they may have contributed to the origin of life on Earth
\citep{Maret2004,Saavik2004,Boogert2008,Herbst2009,Velilla2017}. In dense
molecular clouds, however, photolysis is insufficient, which might otherwise
aid the formation of COMs.  Therefore, it is interesting to study how such
species can be formed without the presence of an embedded energy
source. Formaldehyde is typically detected in the icy mantle covering
interstellar grains and its formation is often attributed to the consecutive
hydrogenation of carbon monoxide (CO) on the surface of amorphous solid water
(ASW) \citep{pet13}.  The further hydrogenation of H$_{2}$CO on the ASW surface will yield
the simplest alcohol, methanol (CH$_{3}$OH) which was repeatedly observed in
dense clouds, both in the solid state and in the gas phase.  The whole process
can be described as~\citep{Tielens1997}
\begin{equation*}
\mathrm{CO} \xrightarrow{\mathrm{H}} \mathrm{HCO} \xrightarrow{\mathrm{H}} 
\mathrm{H_{2}CO} \xrightarrow{\mathrm{H}}  \mathrm{CH_{3}O} \xrightarrow{\mathrm{H}} 
\mathrm{CH_{3}OH} 
\end{equation*}
where only two steps, H + CO and H + H$_{2}$CO, have barriers.  Solid methanol
has been considered as a starting point for the formation of COMs.  To explain
the observed abundance of H$_{2}$CO and CH$_{3}$OH, several surface reactions
need to be investigated and reliable reaction rates should be provided by
experimental and/or theoretical studies.

The formation of formaldehyde and methanol has been well-studied in
experiments.  \citet{Fedoseev2015} experimentally studied surface
hydrogenation of CO molecules under dense molecular cloud conditions. It was
shown that H$_{2}$CO and CH$_3$OH are formed along with molecules with more
than one carbon atom. The latter probably form by recombination of HCO and
CH$_3$O. \citet{Chuang2016} performed laboratory experiments on H-atom
addition and abstraction reactions in mixed CO, H$_{2}$CO and CH$_{3}$OH
ices. They confirmed that H$_{2}$CO, once formed through CO hydrogenation, was
subjected to H-atom addition reactions producing CH$_{3}$OH, but also yielded
CO again via H-atom-induced abstraction reactions.  At the same time,
\citet{Minissale2016} gave their experimental results as H + H$_{2}$CO leading
to two products, CO and CH$_{3}$OH. They found a strong competition between
H-addition (CH$_{3}$OH formation), H-abstraction (CO formation), chemical
desorption of H$_{2}$CO and other surface processes. However, experiments can
only provide possible product channels and the relative amount of products.
It is difficult to obtain the absolute values of rate constants.

A comprehensive theoretical study on the formation of H$_{2}$CO and CH$_{3}$OH
was carried out by \citet{Woon2002}.  He performed quantum chemical electronic
structure calculations for reactions of H + CO and H + H$_{2}$CO in gas, with
water clusters (H$_{2}$O)$_{n}$ ($n\leq 3$) and on water ice.  Without rate
calculations, he qualitatively estimated the catalytic effect of surrounding
water molecules by comparing the changes of activation energies. It was found
that water molecules had little effect on the barrier of the H + CO reaction,
but they modestly enhanced the addition reaction \ce{H + H2CO -> CH3O} and
hindered the abstraction reaction \ce{H + H2CO -> H2 + HCO}.  However, still
no absolute rate constants were given. \citet{Rimola2014} studied the
successive hydrogenation of CO on water ice to finally yield CH$_{3}$OH using
quantum chemical simulations and astrochemical modeling. They compared the
resulting abundances of CO, H$_{2}$CO, and CH$_{3}$OH with respect to water to
those observed in low-mass protostars and dark cores. While they covered the
hydrogenation reactions, they neglected hydrogen abstractions.
 
\citet{Andersson2011} calculated tunneling rate constants of H and D atom
addition to CO at low temperatures.  They provided the unimolecular and
biomolecular rate constants for the H (D) + CO reaction at a temperature range
of 5 to 180~K (20 to 180~K). 

To fully cover the formation of CH$_{3}$OH in the interstellar environment,
the other hydrogenation step with barrier, i.e. H + H$_{2}$CO, is required and
tunneling rate constants should be provided at low temperature.  In this work,
we focus on three possible product channels of H + H$_{2}$CO as follows:
\begin{reaction}
 \ce{H + H2CO -> CH3O}
 \label{re:r1}
\end{reaction}
\begin{reaction}
 \ce{H + H2CO -> CH2OH}
 \label{re:r2}
\end{reaction}
\begin{reaction}
 \ce{H + H2CO -> H2 + HCO}
 \label{re:r3}
\end{reaction}
We study tunneling in those three reactions and provide rate constants for
both, the Eley--Rideal mechanism and the Langmuir--Hinshelwood mechanism on
an ASW surface. In addition, the corresponding deuterium-substituted
reactions are studied and their rate constants are given as well.  Finally, we
provide fits of rate constants to be used in astrochemical models.

\section{Computational Methods} \label{sec:metho}

We calculated the rates for the reactions \ref{re:r1}, \ref{re:r2}, and
\ref{re:r3} on an ASW surface model combining the quantum mechanics/molecular
mechanics (QM/MM) method \citep{war72,war76} and instanton theory
\citep{lan67,mil75,col77,cal77,alt11,ric16}.  The QM/MM method provides the
potential energy, while the semi-classical instanton theory is used for rate
calculations.

\subsection{System Preparation}

We prepared a thermalized ASW surface by classical molecular
dynamics (MD) simulations with NAMD \citep{phi05}. The initial sample was
produced by VMD version 1.9.2~\citep{hum96a} with 9352 TIP3P  \citep{jor83}
water molecules in a slab of 85~{\AA} $\times$ 85~{\AA} and a thickness of
approximately 36~{\AA}.  We used periodic boundary conditions along all three
Cartesian axes and kept a vacuum of around 70~{\AA} thickness between the
slabs. The system was treated in the canonical NVT ensemble. A Langevin
dynamics simulation was performed firstly at 300~K for 100~ps followed by an
instantaneous quenching to 10~K for 20~ps to produce a thermally equilibrated
bulk amorphous solid water at low temperature.  We cut a hemisphere with a
radius of 34~{\AA} to be used in the QM/MM model.

In order to locate different types of binding sites on the ASW surface, the
H$_{2}$CO molecule was placed at 113 initial positions for minimization on a
regular 2D-grid with step size of 2~{\AA} covering the central circle area of
the ASW hemisphere with a radius of 12~{\AA}.  For each of total 113 initial
positions, the H$_{2}$CO molecule was placed 2.5~{\AA} above the ASW surface.
Each H$_{2}$CO molecule and its surrounding water molecules within 6~{\AA}
belonged to QM region (typically about 23 molecules), while the 
water molecules between 6~{\AA} and 12~{\AA} away from the H$_{2}$CO molecule
were in the active MM region (typically about 160 molecules). The other water
molecules, farther than 12~{\AA} from H$_{2}$CO, were frozen.

\subsection{QM/MM Method}

The computational requirements for an accurate quantum mechanical (QM) method
is prohibitive for geometry optimization and tunneling rate calculations on a
system with more than 9000 atoms.  Force field methods are cheap enough
but generally inapplicable for chemical reactions.  In this study, we use a
state-of-the-art hybrid QM/MM simulation which combines the accuracy of the QM
method and the speed of the force field (MM) approach.  We treated the
reactants H, H$_{2}$CO and their immediate environment with density functional
theory (DFT) and described more distant water molecules by the TIP3P force
field \citep{jor83}. Van der Waals parameters for H and H$_{2}$CO were chosen
in analogy to the CHARMM22 force field \citep{mackerell:04} in the QM/MM
treatment.

The hybrid QM/MM simulations were carried out with the
ChemShell~\citep{she03,met14} interface using additive QM/MM coupling.  The
interactions between QM and MM subsystems were handled in the electrostatic
embedding scheme where MM atoms can polarize the electrons in the QM
subsystem. 

The level of theory for the QM region was chosen according to benchmark
calculations performed in the gas phase.  We first optimized the reactant,
transition states and products of the three reactions \ref{re:r1},
\ref{re:r2}, and \ref{re:r3} in the gas phase using the functional
BHLYP-D3~\citep{bec93b,lee88,gri10} and the def2-TZVPD basis set
\citep{rap10}.  Then, the activation energies were calculated using explicitly
correlated unrestricted coupled-cluster with singles and doubles excitations
including perturbative treatment of triple excitations, (U)CCSD(T)-F12,
\citep{adl07,kni09a} based on a restricted Hartree--Fock (RHF) reference
function and the cc-pVTZ-F12 basis \citep{pet08} as reference and with
different DFT funcitonals. The
PWB6K-D3/def2-TZVP~\citep{zhao:05,gri10,florian:05} level resulted in the most
accurate activation energies for all three reactions in comparison with the
(U)CCSD(T) reference as shown in Sec.~\ref{sec:ben}, so this DFT level was
applied in all following QM calculations. Energetic data are given in Kelvin,
rounded to 10~K.

The quantum chemical program package NWCHEM 6.6~\citep{nwchem} was used for
the QM calculations while DL\_POLY~\citep{smith:02} built into ChemShell was
used for the MM region.  The open source DL-FIND~\citep{kaestner:09} optimizer was
used for geometry optimization such as searching for binding sites, transition
states using the dimer method \citep{hen99,kae08}, and instanton pathways with
a modified Newton--Raphson approach \citep{rom11,rom11b}. For the TS searches
and the instanton calculations, the QM part was restricted to the
adsorbate and one water molecule, for instanton searches the active
part as well. 

\subsection{Instanton Theory}

Tunneling rate constants in this work were calculated using instanton
theory \citep{lan67,mil75,col77,cal77,alt11,ric16} based on Feynman
path integral theory in its semiclassical approximation.  In general,
it is only applicable below the crossover temperature
$T_\text{c}$~\citep{gillan:87}:
\begin{equation}
T_\text{c} = \frac{\hbar\omega_\text{b}}{2\pi k_\text{B}}
\end{equation} 
where $\hbar$ is the reduced Planck constant, $\omega_\text{b}$ corresponds to
the absolute value of the imaginary frequency at the transition structure, and $k_\text{B}$ is the
Boltzmann constant.  The semiclassical nature of instanton theory offers a
reasonable ratio of accuracy versus computational cost which is appropriate to
the reactions with small organic molecules on the ASW surface.  At a given
temperature, the instanton itself is the tunneling path with the highest
statistical weight. It can be located using transition state searching methods
implemented in the DL-FIND optimizer.  Integrating along this path and
combining it with the partition function of reactant state, we can compute the
instanton rate constants considering quantum effects such as zero point
vibrational energy (ZPE) and atom tunneling \citep{rom11,rom11b}.  In this
work, for reaction \ref{re:r1} instantons were discretized to 78
images down to 83~K and 154 images below. For reaction \ref{re:r2}, 78 images were used down to 124~K and
154 images below. In reaction \ref{re:r3} we used 78 images down to
111~K and 154 images below that. Branching ratios $R$ were calculated by
$R_i=k_i/(k_1+k_2+k_3)$ from the rate constants $k_i$.

\section{Results and Discussion \label{sec:resul}}

\subsection{Benchmark Calculations\label{sec:ben}}

Benchmark calculations were performed to identify an accurate DFT level for
binding site optimization, transition state search and instanton rate
calculations.  We applied the BHLYP-D3/def2-TZVPD level for the optimization
of the reactant, transition states and products in the gas phase and used the
resulting geometries to calculate the activation energies for three product
channels on the (U)CCSD(T)-F12/cc-pVTZ-F12 level and with various DFT
functionals with two different basis sets, def2-TZVP~\citep{florian:05} and
def2-SVPD~\citep{rappoport:10}.  The resulting activation energies are listed
in Table~\ref{tab:tab1}.  Compared with the (U)CCSD(T)-F12/cc-pVTZ-F12 result
as the reference, the functional PWB6K-D3 (PWB6K functional with D3 dispersion
correction) with the def2-TZVP basis set gave the closest activation energies
for all three product channels with a relative deviation smaller than 7\%.
The results from other functionals had much larger deviations, some of which
with opposite signs are qualitatively wrong, such as PBE96-D3, SSB-D-D3,
TPSS-D3 and TPSSH-D3.  We chose PWB6K-D3/def2-TZVP as the proper DFT level and
re-optimized all species at this level.  The activation energies at
(U)CCSD(T)-F12/cc-pVTZ-F12 and PWB6K-D3/def2-TZVP levels were re-calculated
and listed at the last two lines in Table~\ref{tab:tab1}.  The different
geometries resulted in slightly different activation energies. However,
PWB6K-D3/def2-TZVP still gave good results (within a maximum of 220~K
deviation, 9\% relative deviation) compared to the reference
UCCSD(T)-F12/cc-pVTZ-F12 data.

\subsection{H$_2$CO Binding Sites and Binding Energies}

\begin{figure}[htbp]
\includegraphics[width=8cm]{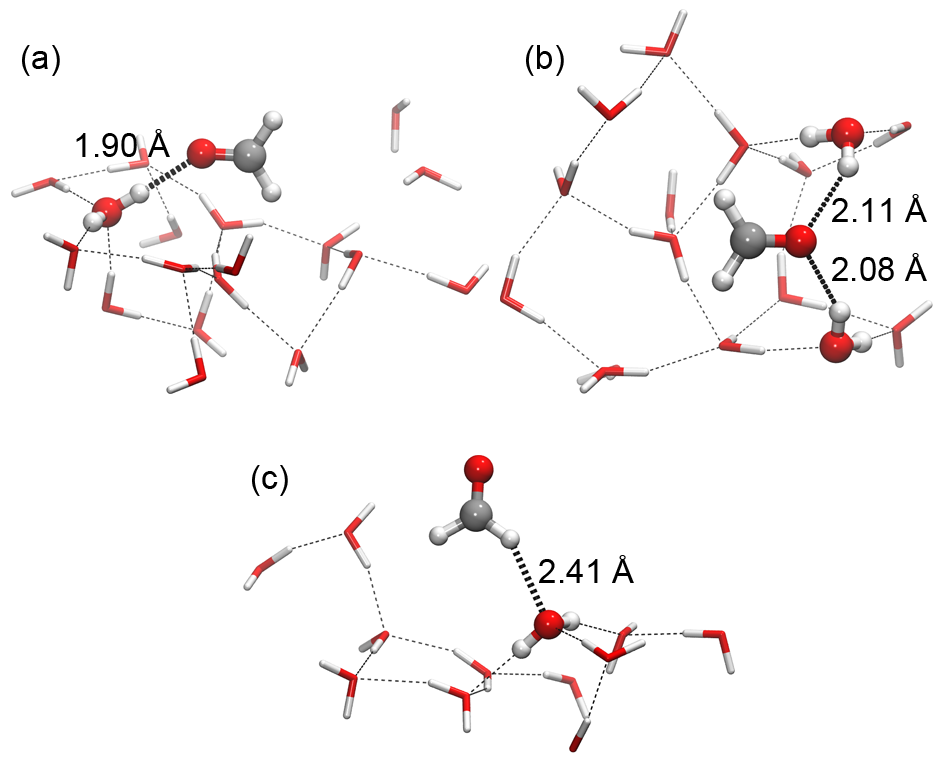}
\caption{Three different H$_{2}$CO binding types on the amorphous solid water
  surface. Only QM molecules are shown. The H$_{2}$CO molecule and water
  molecules directly connected via hydrogen bonds are shown as balls and sticks.
  Bond lengths are given in \AA. \label{fig:bs}}
\end{figure}



All three reactions originate from H$_2$CO bound to the ASW surface. Thus, we
first investigated possible binding sites and associated binding modes and
energies.  The H$_{2}$CO molecule was initially placed 2.5~{\AA} above the ASW
surface. Local minimum searches were started from 113 H$_{2}$CO initial
positions. Out of these, 81 cases converged to physically meaningful local
minima. They are categorized in three different types of binding modes,
visualized in Fig.~\ref{fig:bs}.  In this figure, we only show the molecules
in the QM region, while the calculations were performed for the full QM/MM
model.  Type~(a) is the major binding mode, 89\% of the cases were found in
that mode. The H$_{2}$CO accepts one hydrogen bond at its O atom.  The bond
lengths of the hydrogen bonds vary from 1.80~{\AA} to 2.50~{\AA} with an
average of 1.97~{\AA}. There were 7 cases (9\%) belonging to type~(b). In this
case, the O end of the H$_{2}$CO molecule accepted two hydrogen bonds from two
different water molecules.  The remaining two cases (2\%) belonged to
type~(c). Here, H$_{2}$CO donates a hydrogen bond to an O atom of water. These
rather weak hydrogen bonds had lengths of 2.41~{\AA} and 2.89~{\AA}.

\begin{figure}[htbp]
\begin{center}
\includegraphics[width=8cm]{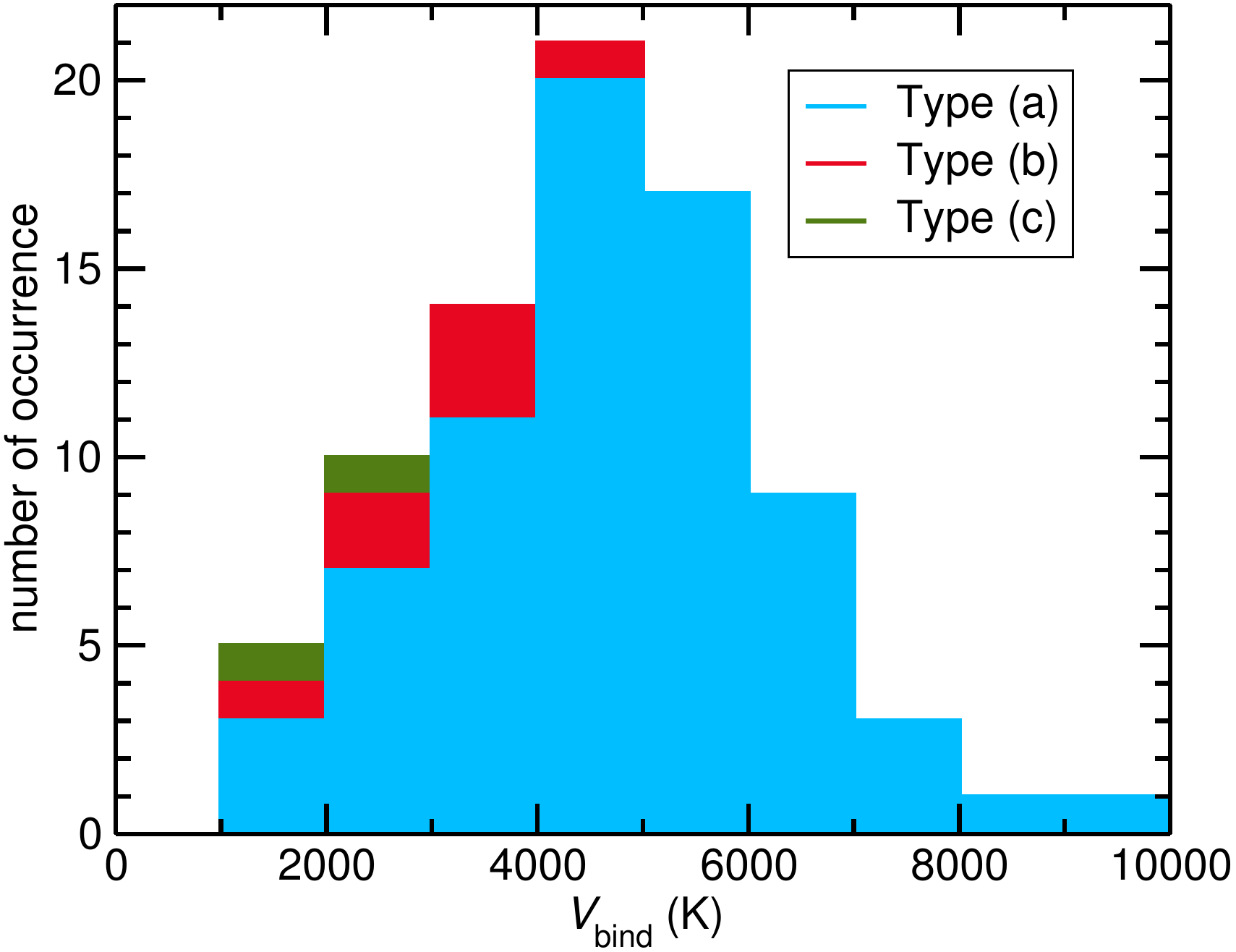}
\end{center}
\caption{The distribution of H$_{2}$CO binding energies on the amorphous solid
  water surface at the PWB6K-D3/def2-TZVP/MM level of theory.\label{fig:be}}
\end{figure}

The binding energy $V_\text{bind}$ of H$_{2}$CO is the potential
energy required to disassemble the H$_{2}$CO molecule from the ASW
surface to the gas phase including the relaxation of the surface.
Figure~\ref{fig:be} illustrates the binding energy distribution of
H$_{2}$CO molecules on the ASW surface using the 81 successful
cases. The distribution of binding energies corresponding to binding
site type~(a) is broad, ranging from 1000 to 9370~K with the largest
fraction between 4000 and 5000~K. $V_\text{bind}$
for all sites of type~(a) is larger than 1200~K.  There were only 9
cases of binding energies belonging to type~(b) and type~(c). They are
rather weakly bound, with binding energies below 5000~K. These binding
energies were obtained without consideration of the zero point
vibrational energy (ZPE). We calculated the ZPE contributions for 5
cases. Taking ZPE into account reduces the binding energy by 580 to
920~K.

\subsection{H$_{2}$CO + H Transition States}

\begin{deluxetable}{llrcrrrr}
\tablenum{2}
\tablecaption{Comparison of transition states for reactions~\ref{re:r1}, 
\ref{re:r2} and \ref{re:r3} in gas and on the amorphous solid water surface. 
Energies (ER process) are given in K, bond distances $d$ in \AA, 
frequencies in cm$^{-1}$ and the temperature in K. \label{tab:tab2}}
\tablewidth{0pt}
\tablehead{
\colhead{site}        & \colhead{reaction}      & \colhead{$V_\text{bind}$}         
& \colhead{$d$} & \colhead{$\omega_\text{b}$} & 
\colhead{$V_\text{act}$} & \colhead{$E_\text{act}$} & \colhead{$T_\text{c}$} 
}
\startdata
{     }  &  \ref{re:r1}  &      & 1.855 &  840  &1620  & 2160 & 192 \\
 gas     &  \ref{re:r2}  &      & 1.515 & 1521  &4780  & 5210 & 348 \\
{     }  &  \ref{re:r3}  &      & 1.043 & 1599  &3320  & 2470 & 366 \\
\hline                                                       
{     }  &  \ref{re:r1}  &      & 1.848 &  831  &1420  & 1900 & 190 \\
 ASW 1   &  \ref{re:r2}  & 4600 & 1.488 & 1637  &5220  & 5670 & 375 \\
{     }  &  \ref{re:r3}  &      & 1.022 & 1663  &4030  & 3030 & 381 \\
\hline                                                       
{     }  &  \ref{re:r1}  &      & 1.853 &  793  &1470  & 1920 & 182 \\
 ASW 2   &  \ref{re:r2}  & 5420 & 1.490 & 1626  &5200  & 5620 & 372 \\
{     }  &  \ref{re:r3}  &      & 1.034 & 1567  &3200  & 2270 & 359 \\
\hline                                                       
{     }  &  \ref{re:r1}  &      & 1.848 &  832  &1410  & 1890 & 191 \\
 ASW 3   &  \ref{re:r2}  & 4270 & 1.515 & 1525  &6480  & 6670 & 349 \\
{     }  &  \ref{re:r3}  &      & 1.044 & 1453  &4590  & 3640 & 333 \\
\hline                                                       
{     }  &  \ref{re:r1}  &      & 1.854 &  819  &1380  & 1870 & 188 \\
 ASW 4   &  \ref{re:r2}  & 5580 & 1.491 & 1619  &5780  & 6090 & 371 \\
{     }  &  \ref{re:r3}  &      & 1.021 & 1665  &4850  & 3680 & 381 \\
\hline                                                       
{     }  &  \ref{re:r1}  &      & 1.853 &  813  &1100  & 1640 & 186 \\
 ASW 5   &  \ref{re:r2}  & 6080 & 1.507 & 1544  &5190  & 5620 & 354 \\
{     }  &  \ref{re:r3}  &      & 1.034 & 1598  &3920  & 2970 & 366 \\
\enddata
\end{deluxetable}

We investigated the H + H$_{2}$CO transition states for the three product
channels both in gas phase and on the ASW surface.  From the dominant binding
type~(a) as shown in the Fig.~\ref{fig:bs}, we selected 5 binding sites with
binding energies between 4200 and 6100~K to perform transition state
optimizations using the QM/MM model.  For each binding site, we found
transition states TS1, TS2 and TS3 corresponding to reaction channels
\ref{re:r1}, \ref{re:r2}, and \ref{re:r3}, respectively. The results are given
in Table~\ref{tab:tab2}.  The label ASW~$n$ ($n=1-5$) is used to distinguish
transition states based on different binding sites. The distance $d$
  measures the distance from the incomming H atom, specifically the C--H
  distance for \ref{re:r1}, the O--H distance for \ref{re:r2}, and the H--H
  distance for \ref{re:r3}.  For comparison, the transition states of the gas
phase reactions are provided in this table as well. The label $V_\text{act}$
refers to the activation energy as potential energy difference, the label
$E_\text{act}$ indicates that the difference in vibrational zero point energy
was taken into account as well.

\begin{figure}[htbp]
\plotone{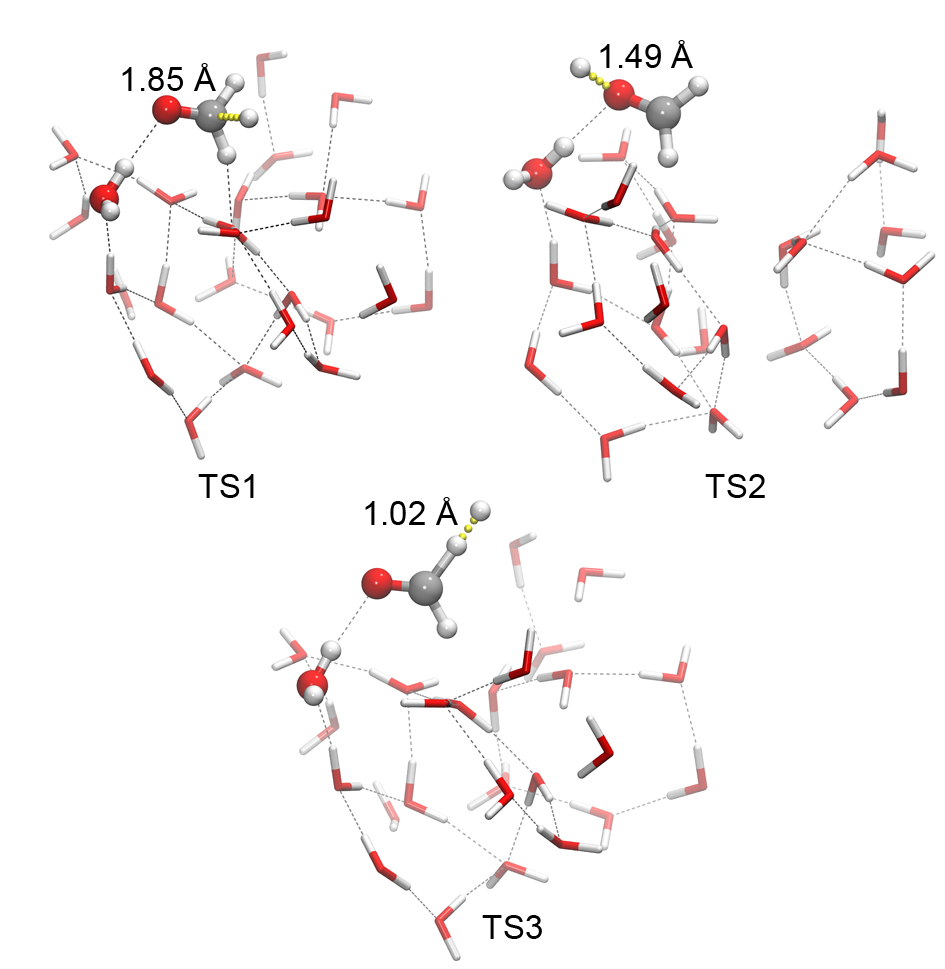}
\caption{Optimized geometries of the transition states TS1, TS2 and TS3 for
  the reactions H +
  H$_{2}$CO on the amorphous solid water surface ASW~1 yielding CH$_{3}$O,
  CH$_{2}$OH and H$_{2}$ + HCO.  In the TS search, the ball-and-stick
  molecules were in the QM region, while others are in the MM active region.
  Forming bonds are indicated by yellow dots. \label{fig:ts}}
\end{figure}

From a classical perspective, H addition to C, \ref{re:r1}, has the lowest
activation energies and is expected to be the dominant reaction channel while
the H addition to O, \ref{re:r2}, has the highest activation energies and will
happen in negligible amounts only.  The H abstraction reaction, \ref{re:r3},
has a medium barrier height. Transition structures are shown in
Fig.~\ref{fig:ts}.

Compared with transition state geometries in the gas phase, the ones on the
ASW surface always have shorter bond lengths along the reaction coordinates
(H--C for TS1, H--O for TS2, and H--H for TS3, see Fig.~\ref{fig:ts}),
i.e. are later transition states. The exception is ASW~3 for TS2 and TS3,
which are very slightly earlier transition states than the gas-phase ones. The
effect of the ASW surface is to lower the barrier for \ref{re:r1} and increase
the barriers for \ref{re:r2} and \ref{re:r3}. Thus, it promotes the formation
of the more complex molecules CH$_{3}$O and CH$_{3}$OH and hinders the
destruction of H$_{2}$CO to HCO. The reason is probably that the more
protonated species CH$_{3}$O and CH$_{3}$OH are also more polar than HCO and
therefore form more stable hydrogen bonds to the water environment.

All three reactions are exothermic, the most stable product is
CH$_2$OH. Including ZPE, the reaction energy of \ref{re:r1} is $-12,510$~K in
the gas phase and $-12,160$~K on ASW in the binding site ASW~1. For
\ref{re:r2} the values are $-14,980$~K and $-14,660$~K. Reaction \ref{re:r3}
is much less endothermic with a gas-phase reaction energy of $-7430$~K and a
surface reaction energy (with H$_2$ desorbed into the gas phase) of $-6220$~K.
Thus, under thermodynamic control, the main product of hydrogenation of
H$_2$CO would be CH$_2$OH.

\subsection{Tunneling Rate Constants}


\begin{figure*}
\gridline{\fig{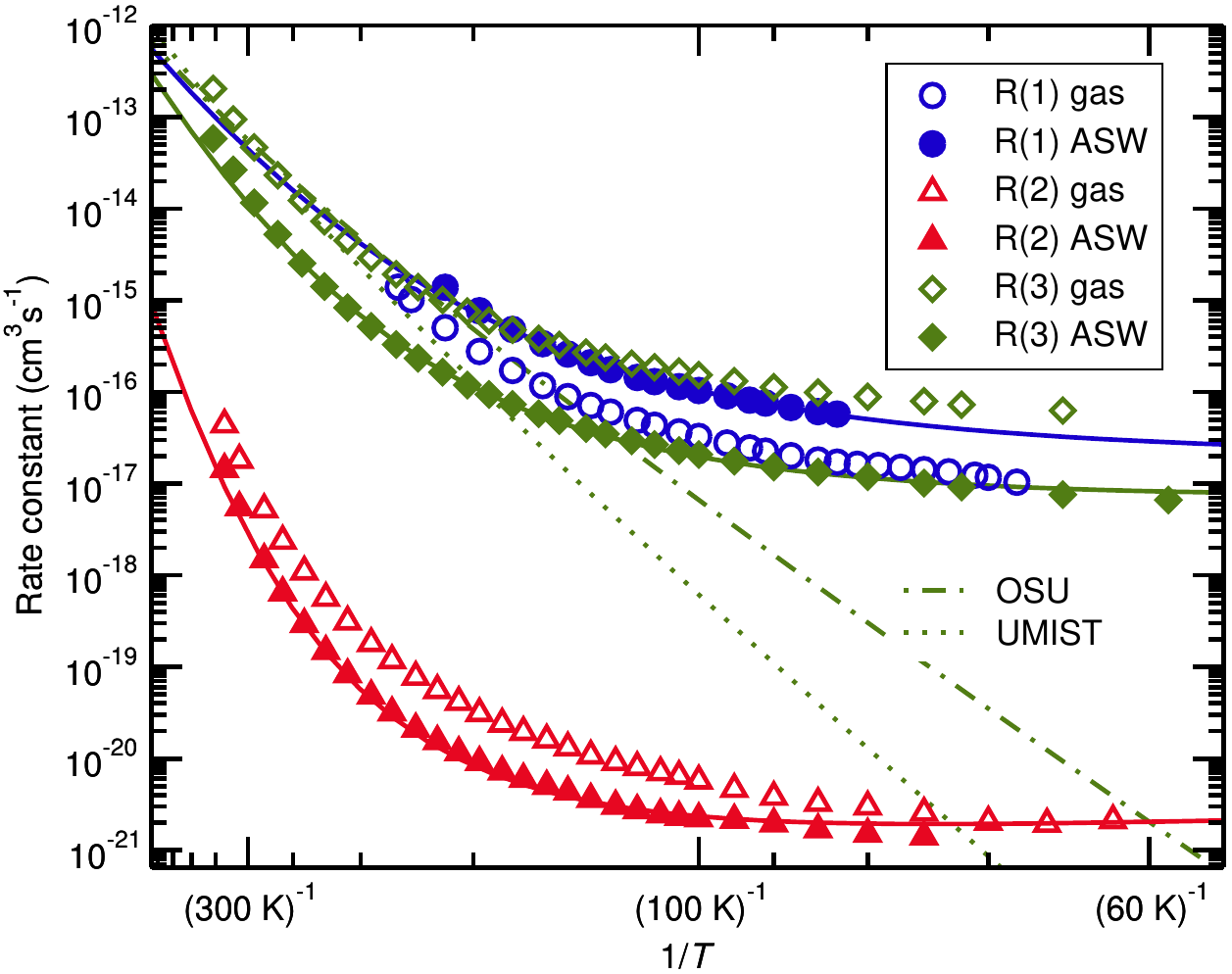}{8cm}{(a) Eley--Rideal mechanism}
          \fig{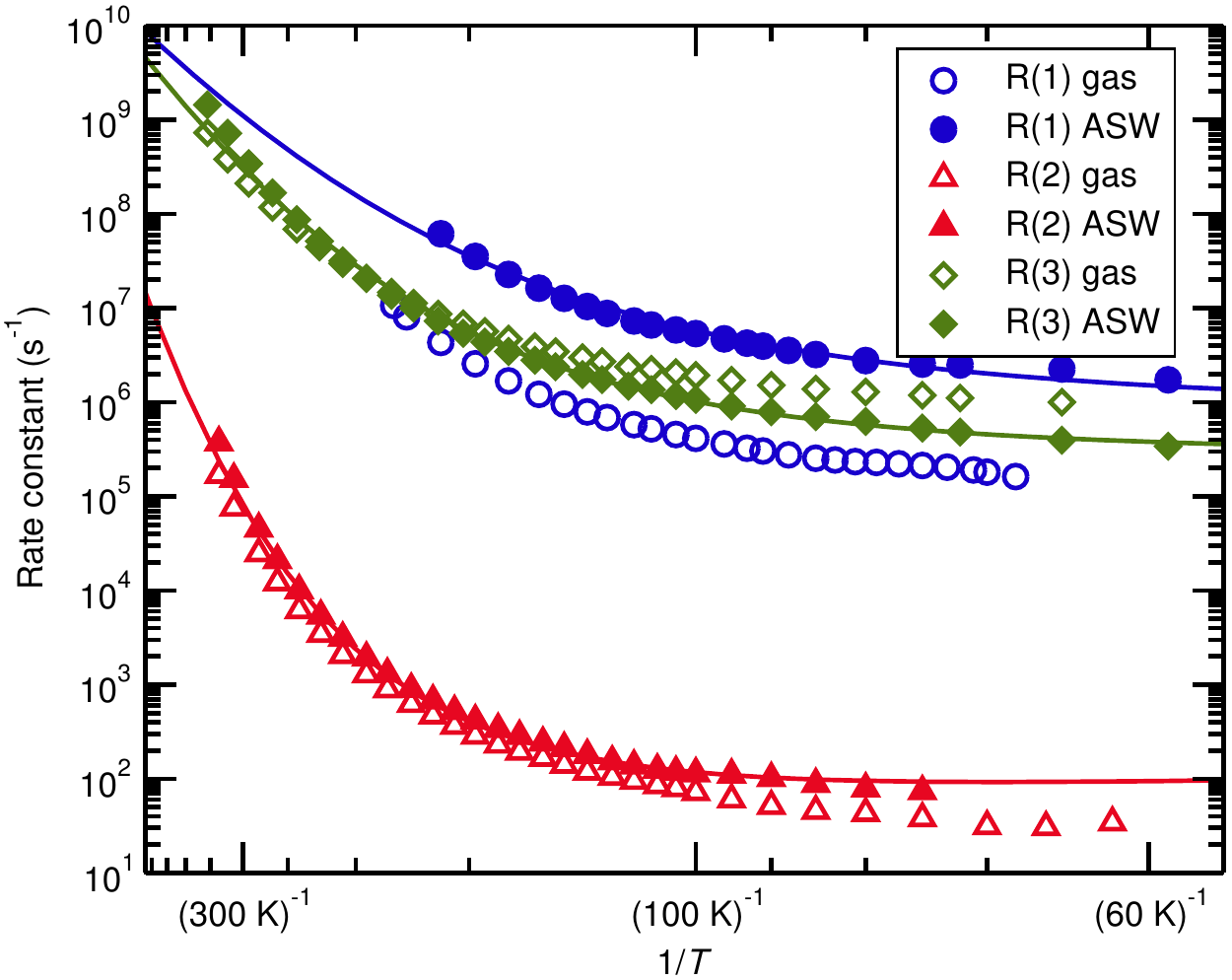}{8cm}{(b) Langmuir--Hinshelwood mechanism}}
\caption{Instanton rate constants labeled as different markers and fitted
  lines for the reactions of H + H$_{2}$CO yielding CH$_{3}$O (\ref{re:r1}),
  CH$_{2}$OH (\ref{re:r2}), and H$_{2}$ + HCO (\ref{re:r3}) in gas and on the
  amorphous solid water surface. OSU and UMIST gas-phase data for \ref{re:r3} are shown for
  comparison.\label{fig:instant}}
\end{figure*}


Starting from transition states TS1, TS2 and TS3, we calculated tunneling rate
constants for reactions~\ref{re:r1}, \ref{re:r2}, and \ref{re:r3} both in the
gas phase and on the ASW surface using instanton theory. From
Table~\ref{tab:tab2} it is clear that the difference between the different
adsorption sites is small. Therefore, we restricted the following analysis to
site ASW~1. We provide surface instanton tunneling rate constants from just
below the individual crossover temperatures down to 59~K, 75~K, and 59~K for
the three product channels in supplementary machine-readable files.  In the
case of the gas-phase reaction, the lowest temperatures are 68~K, 62~K and
65~K. Below that, the tunneling energy dropped below the energy of the
reactants and bimolecular rate constants are not accessible any more.  For the
surface reactions, we considered both the Eley--Rideal mechanism (ER) and the
Langmuir--Hinshelwood mechanism (LH). The results are shown in
Fig.~\ref{fig:instant}, where they are also compared to UMIST and OSU
parameterizations for \ref{re:r3}. We used hollow and solid markers to
distinguish gas phase and surface reactions and labeled blue circles, red
triangles and green diamonds for reactions~\ref{re:r1}, \ref{re:r2}, and
\ref{re:r3} separately.  The markers designate the gas phase and surface
instanton rate constants while the lines correspond to the fits described in
detail in Sec.~\ref{sec:fit}.

Fig.~\ref{fig:instant} shows the rate constants and the results of the
fits. The significant flattening of the Arrhenius curves indicates strong
quantum tunneling effects in all  three reactions.  It is
obvious that reaction~\ref{re:r2}, forming CH$_{2}$OH, is the slowest reaction
both in the gas phase and on the ASW surface owing to its high barriers of
5210~K and $\sim$6000~K for those two situations. The rate constant of
\ref{re:r2} is about four orders of magnitude lower than those of
\ref{re:r1} and \ref{re:r3} at low temperature.

The rate constants of \ref{re:r1} and \ref{re:r3} are similar in
magnitude. In the
gas phase, the hydrogen abstraction reaction~\ref{re:r3}, forming H$_{2}$ +
HCO has larger rate constants than the hydrogen addition reaction~\ref{re:r1},
forming CH$_{3}$O, at temperatures below 180~K. This is due to the stronger
tunneling effects in \ref{re:r3}. The activation energy of \ref{re:r3} is with
2470~K slightly higher than that of \ref{re:r1} with 2160~K, see
Table~\ref{tab:tab2}. Thus, at high temperature, when tunneling is suppressed,
\ref{re:r1} is faster. It should be noted here, that the hydrogen additions,
\ref{re:r1} and \ref{re:r2} can, actually, only happen on the surface because
of the limited efficiency of dissipation of the reaction energy in the gas
phase.

\begin{figure*}
\begin{center}
\includegraphics[width=8cm]{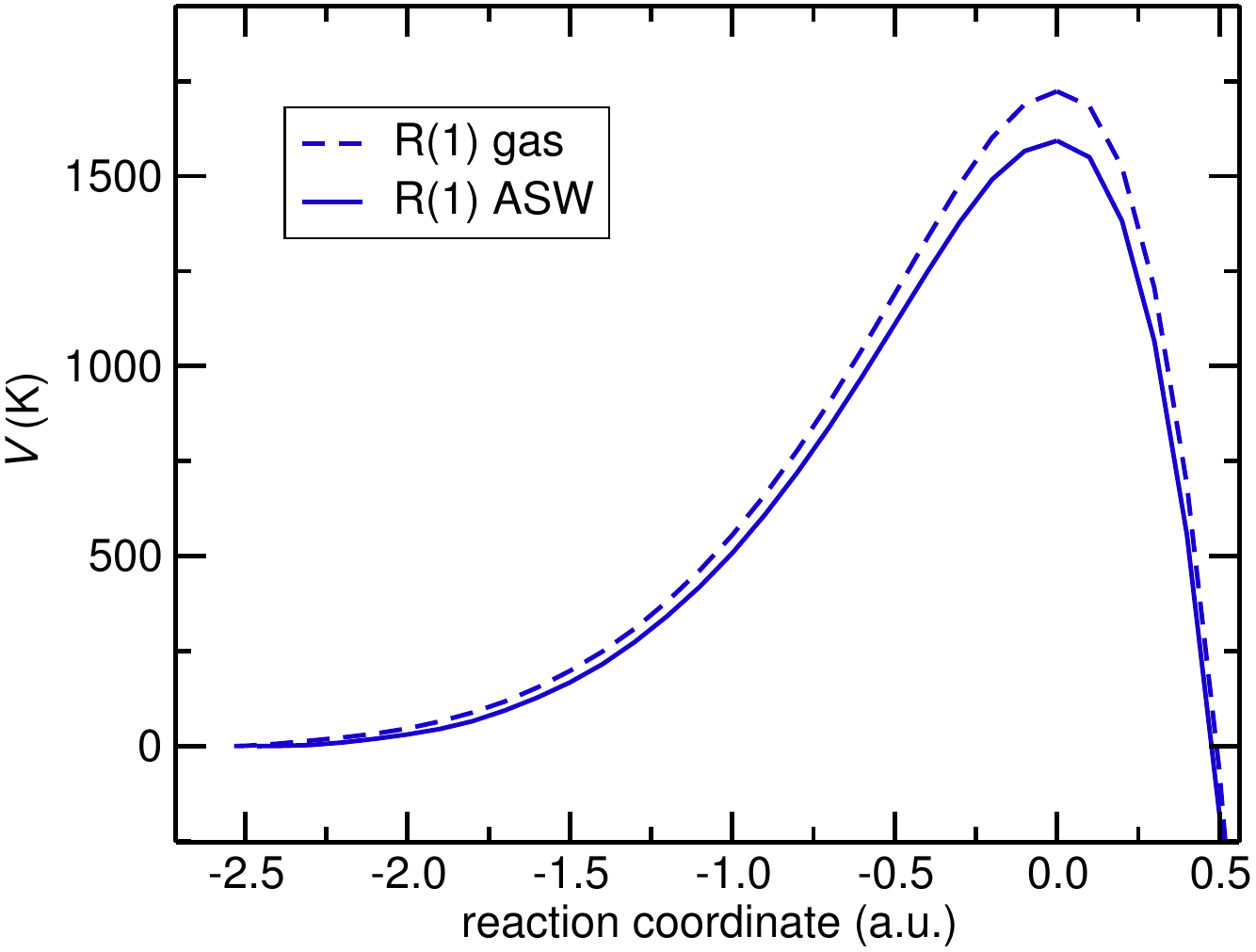}\ \ \
\includegraphics[width=8cm]{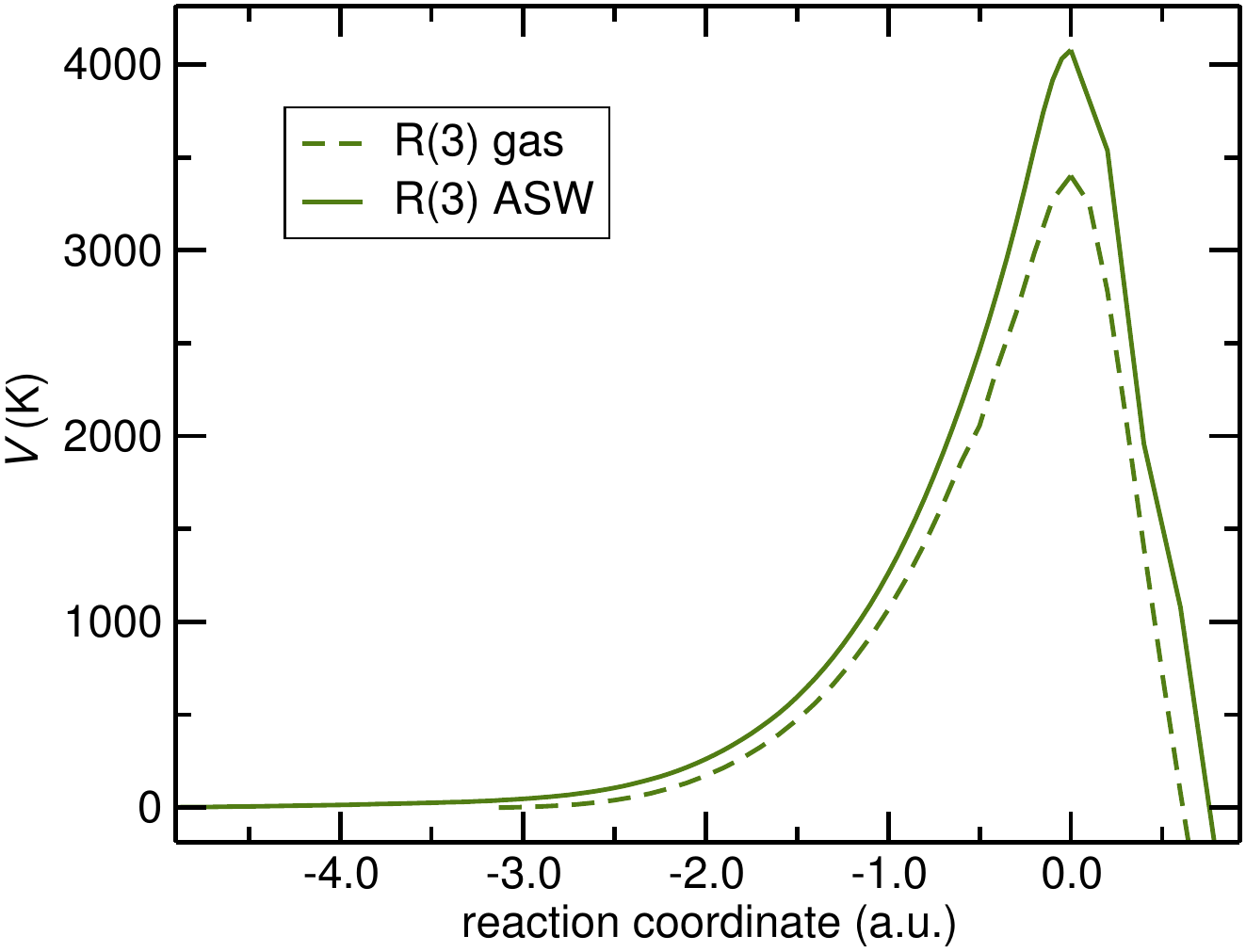}
\end{center}
\caption{The minimum energy paths of the reactions of \ce{H + H2CO -> CH3O}
  and \ce{H + H2CO -> H2 + HCO} in the gas phase and on the amorphous solid water
  surface.\label{fig:irc}}
\end{figure*}

On the ASW surface, the situation between \ref{re:r1} and \ref{re:r3} is
reversed. The reaction~\ref{re:r1}, forming CH$_{3}$O, is the preferred
channel compared to hydrogen abstraction \ref{re:r3} because of catalytic
effects of the ASW surface.  As shown in Table~\ref{tab:tab2}, the ASW surface
decreased the bimolecular activation energy of \ref{re:r1} from 2160~K to
1900~K (for binding site ASW~1, the values are similar for the other sites),
while it raised the activation energy of reaction~\ref{re:r3} from 2470~K to
3030~K. That is to say, the ASW surface increased the difference in barrier
heights between \ref{re:r1} and \ref{re:r3} from 310~K to 1130~K. To explain
that, we calculated the intrinsic reaction coordinate (IRC) to show the
barrier shapes for \ref{re:r1} and \ref{re:r3} both in gas and on the ASW. The
IRC data are shown in Fig.~\ref{fig:irc}. Note that these are potential
energies and do not include ZPE. The ASW-barrier of \ref{re:r1} is lower, but
a little bit broader than the gas-phase one, while the ASW-barrier of
\ref{re:r3} is higher and narrower than that of the gas phase.  It
demonstrates that the ASW catalytic influence on the barrier height increases
the rate of \ref{re:r1} while its influence on the barrier width increases
\ref{re:r3}. Including tunneling at the temperatures considered here, the
effect on the height dominates, \ref{re:r1} is the most effective reaction
channel. The ASW surface promotes the formation of the radical CH$_{3}$O at
the low temperatures of the interstellar medium, which is expected to increase
the the yield of methanol and other more complex radicals and molecules. The
reason, again, is probably that the polar water environment stabilizes the
polar CH$_{3}$O (gas-phase dipole moment of 2.0 Debye) more efficiently than
the comparatively less polar HCO (1.7 Debye).

It may be argued that the hydrogenation reactions of CO are more likely to
happen on a CO ice surface than on ASW. In dense clouds, however, one can
expect even a CO surface to already contain a significant part of impurities
bearing OH groups, like methanol. Such an environment is expected to have a
similar effect on the reactions studied here as ASW.

\begin{figure}[htbp]
\begin{center}
\includegraphics[width=8cm]{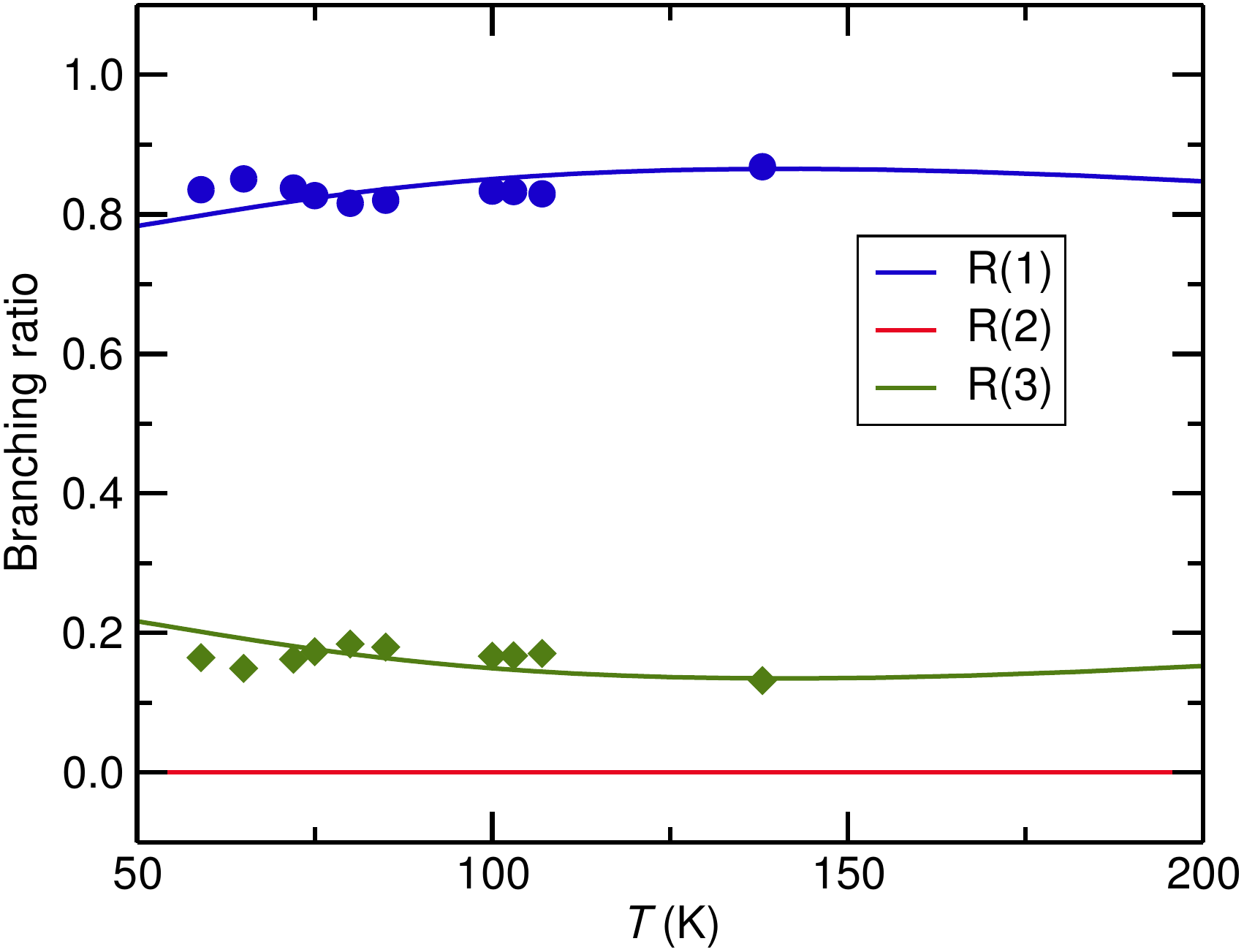}
\end{center}
\caption{Branching ratios for the three reaction channels of the LH
  mechanism on an ASW surface. Lines correspond to the fits, symbols to
  ratios of instanton rate constants.\label{fig:br}}
\end{figure} 

In panel~(a) of Fig.~\ref{fig:instant}, we show the bimolecular instanton rate
constants. These data represent the ER mechanism, in which one reactant
(H$_{2}$CO) is adsorbed on the surface and reacts with a second reactant (H
atom) that approaches directly from the gas phase, related to a biomolecular
reaction between the adsorbate-surface system and the incoming atom.

Unimolcular rate constants, which relate to the LH mechanism, are shown in
panel~(b) of Fig.~\ref{fig:instant}. The LH mechanism is expected to dominate
on the ASW surface at low temperature. In that case, an encounter complex of H
with H$_{2}$CO on the ASW surface reacts to CH$_{3}$O, CH$_{2}$OH or H$_{2}$
+ HCO bound on the ASW surface in the three product channels.  The activation
energies $E_\text{act}$ are 1890~K, 5660~K, and 3030~K in those three reactions,
respectively.  The unimolecular rate constants behave similarly to the
bimolecular ones. Again, ASW leads to a preference of \ref{re:r1} over
\ref{re:r3} in contrast to the gas phase data. \ref{re:r2} is also negligible
for the unimolecular case. 

In Fig.~\ref{fig:br} we show the branching ratios for the three reactions for
the LH mechanism on an ASW surface. At $T=200$~K, \ref{re:r1} accounts for
about 85\% of the reactions, \ref{re:r3} for 15\% and \ref{re:r2} for
0.001\%. At $T=50$~K the fraction of \ref{re:r3} increases a bit to
22\%. Thus, it is clear that the formation of CH$_{3}$O is the dominant
channel resulting from hydrogenation of H$_{2}$CO. Hydrogen abstraction,
\ref{re:r3}, happens with a significant fraction as well, however. The neglect
of \ref{re:r3} by \citet{Rimola2014} may be the reason for the somewhat high
fraction of methanol they found compared to observations.

\subsection{Kinetic Isotope Effects}

\begin{figure*}
\begin{center}
\includegraphics[width=5.8cm]{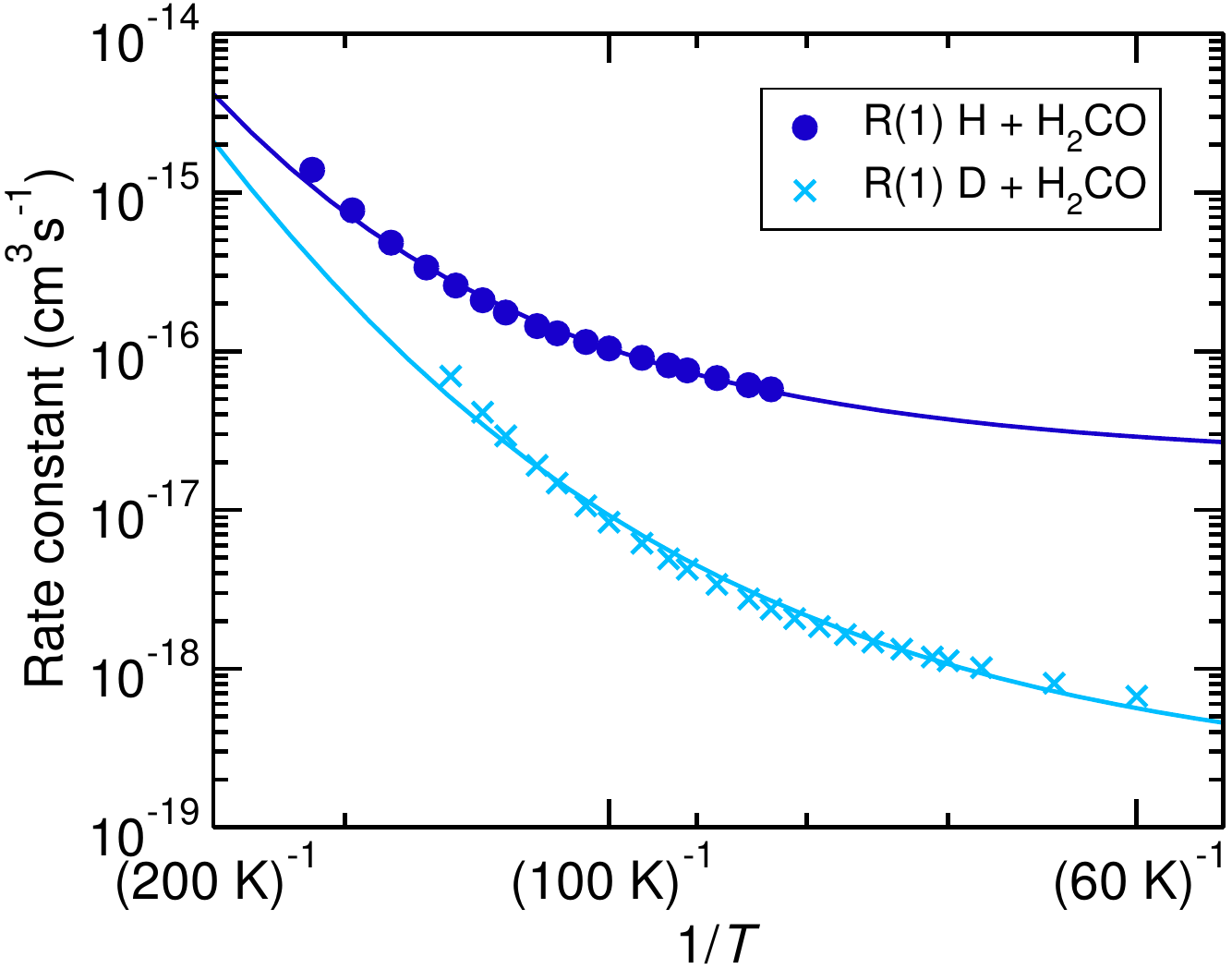}\ \ 
\includegraphics[width=5.8cm]{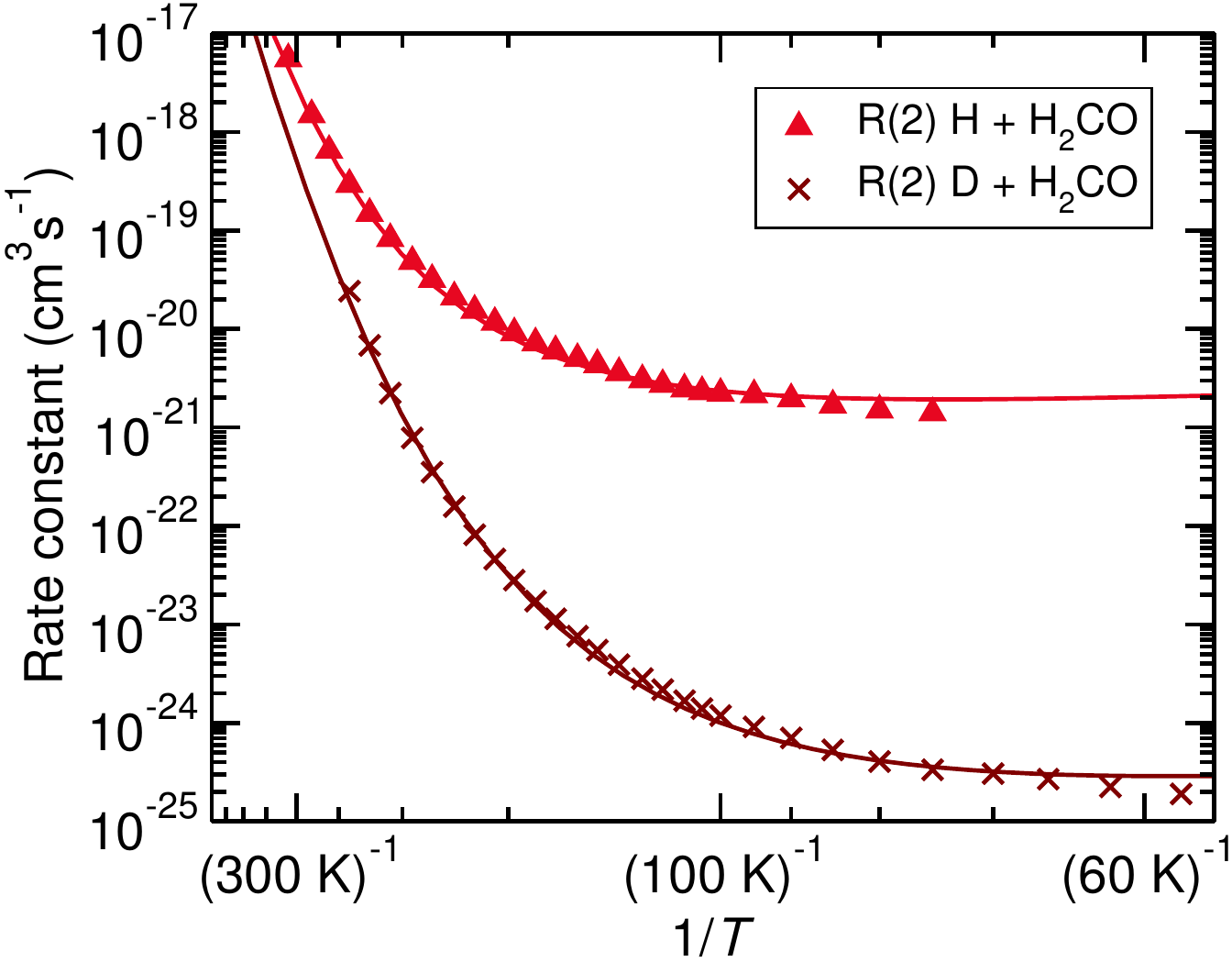}\ \ 
\includegraphics[width=5.8cm]{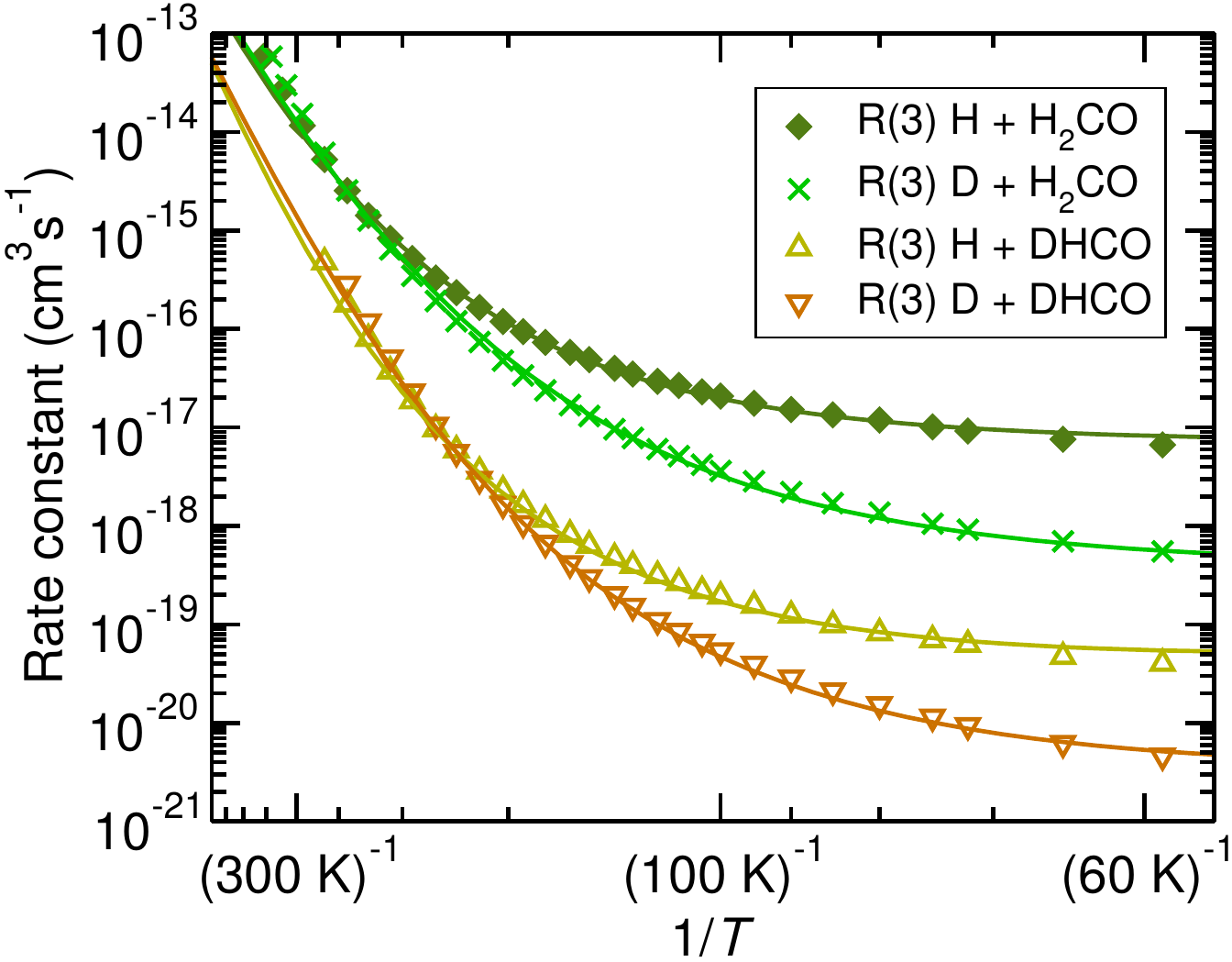}\\
(a) Eley--Rideal mechanism\\
\includegraphics[width=5.8cm]{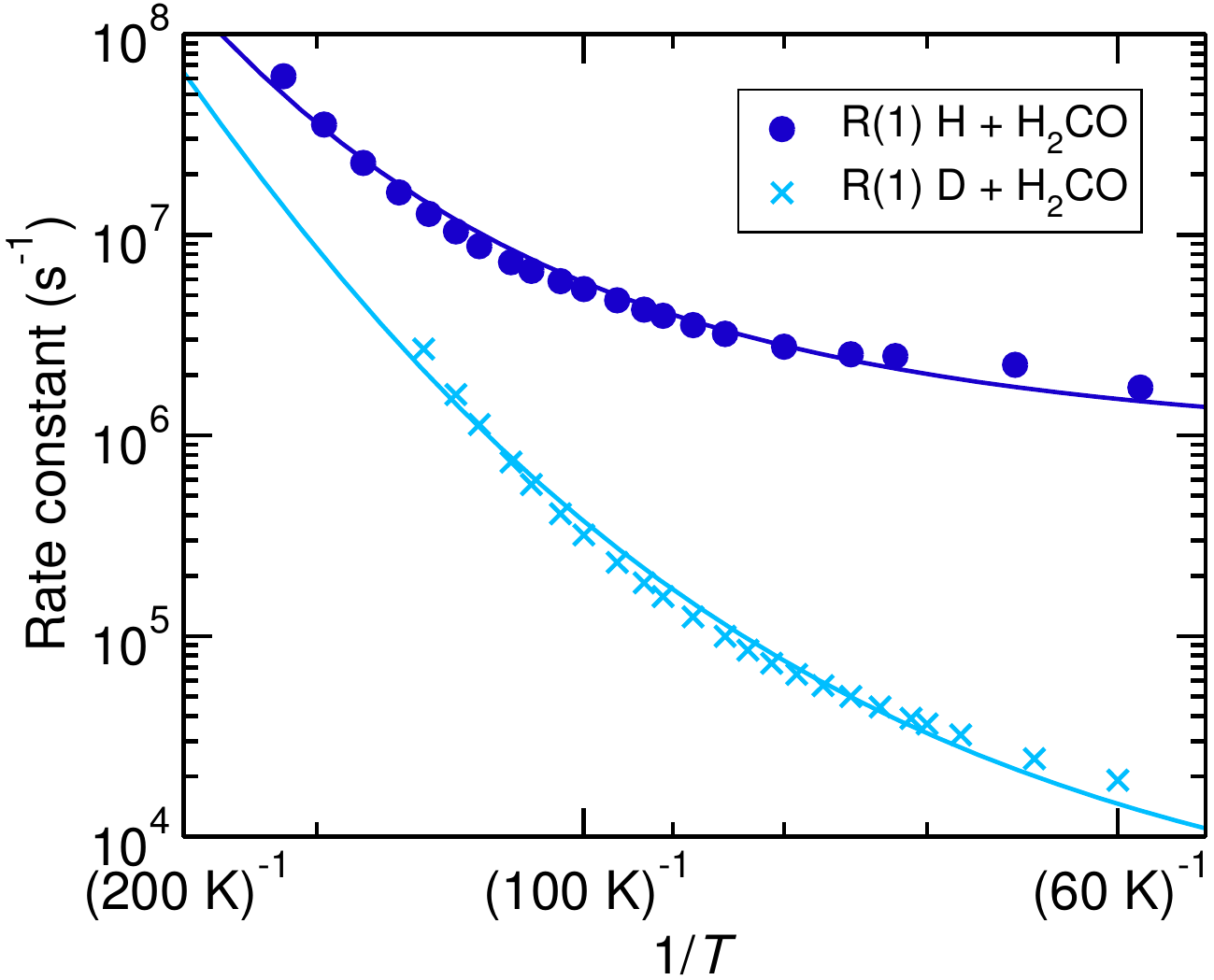}\ \ 
\includegraphics[width=5.8cm]{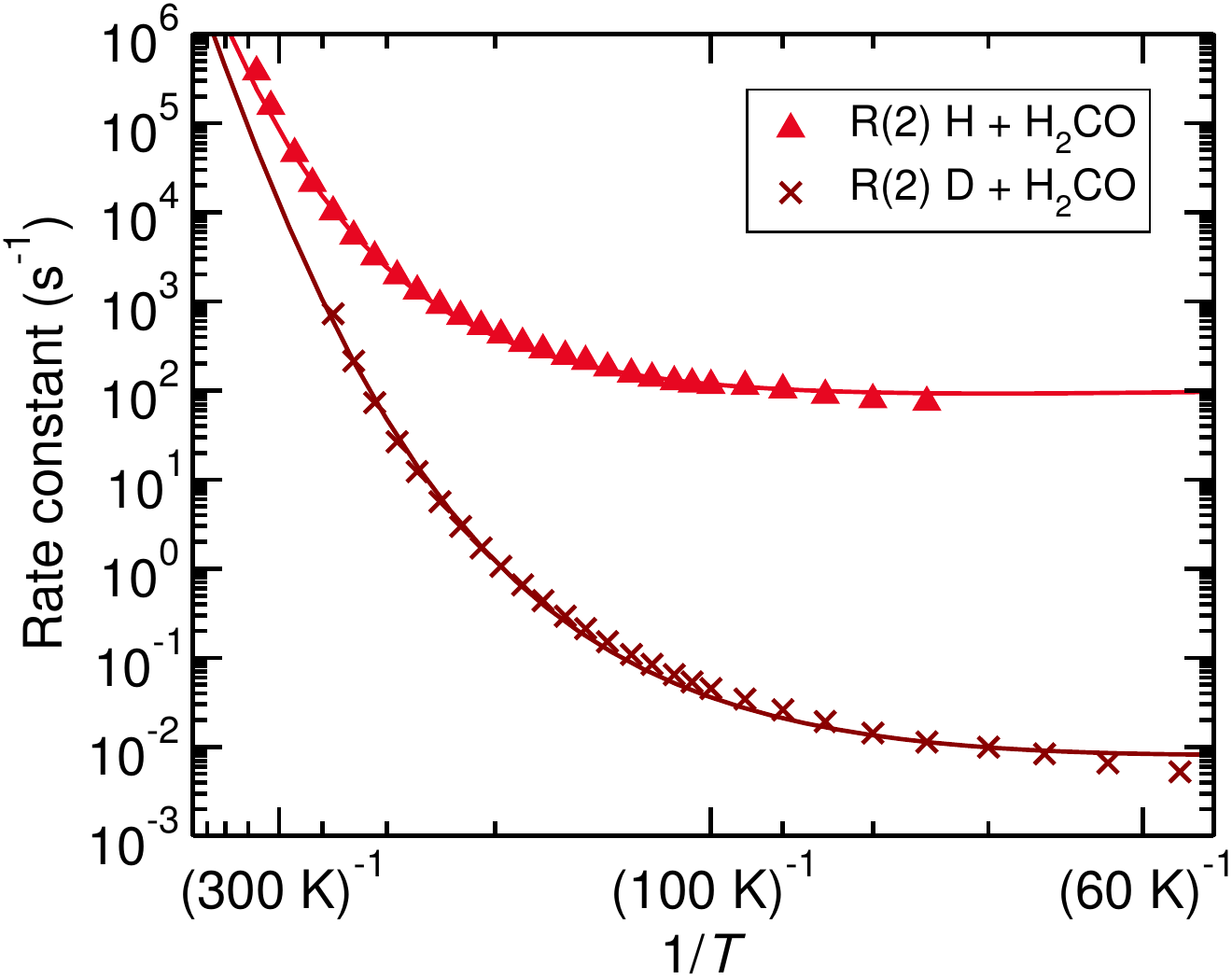}\ \ 
\includegraphics[width=5.8cm]{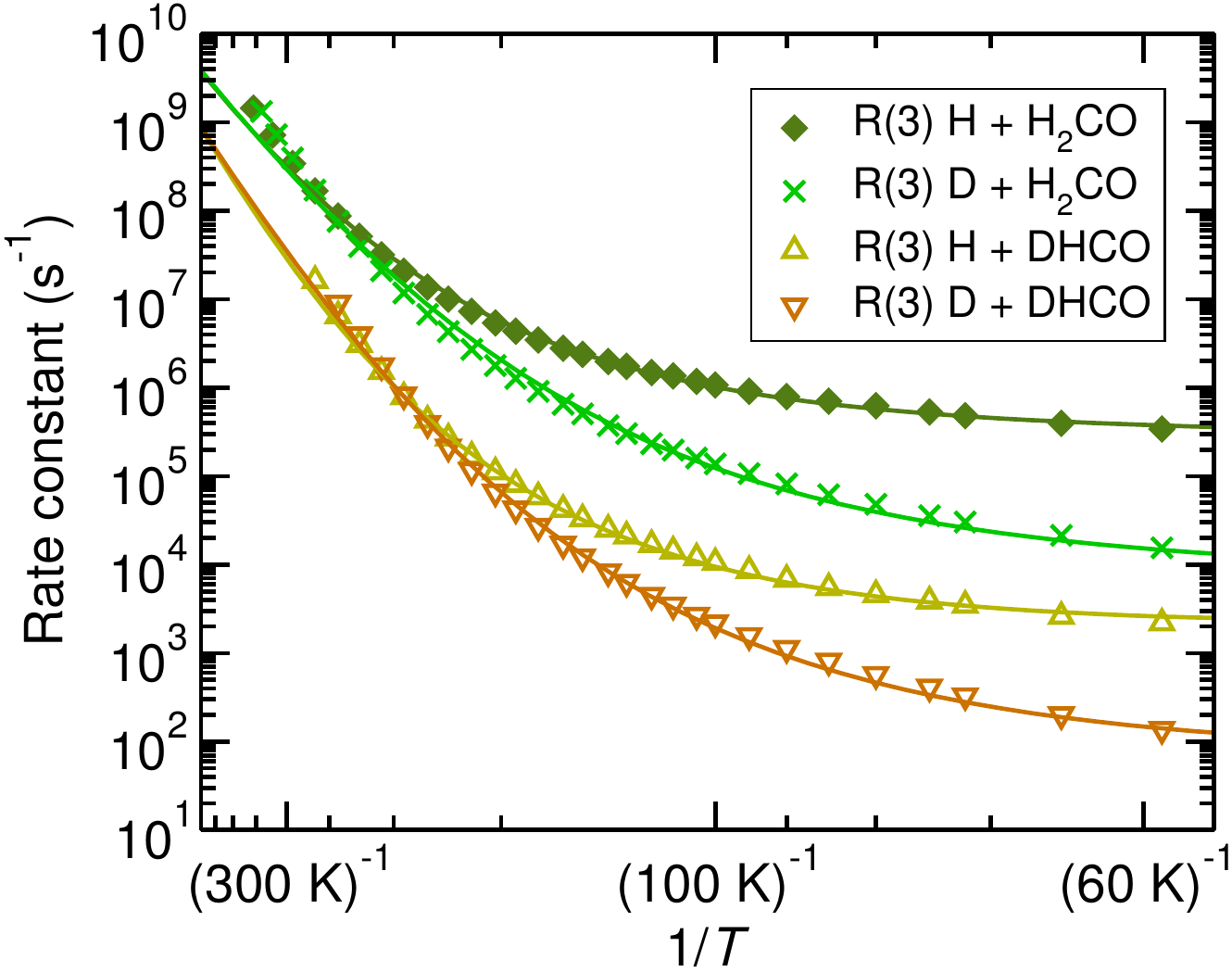}\\
(b) Langmuir--Hinshelwood mechanism
\end{center}
\caption{Kinetic isotope effects of the H+H$_{2}$CO reactions on the 
amorphous solid water surface yielding CH$_{3}$O, CH$_{2}$OH and H$_{2}$ + HCO. 
The markers stand for the calculated instanton rate constants while the 
lines are the fitted results. \label{fig:kie}}
\end{figure*}

Owing to the importance of the deuterium chemistry for tracing dynamic
properties of a cloud, we investigated the H/D kinetic isotopic effects (KIEs)
for the reactions~\ref{re:r1}, \ref{re:r2} and \ref{re:r3}. The KIE is defined
as the ratio of rate constants for the reactions involving the light and the
heavy isotopologues, i.e. KIE = $k_\text{H}/k_\text{D}$ where $k_\text{H}$ and
$k_\text{D}$ are the rate constants for the reactions with protium and
deuterium.  Both the unimolecular and the biomolecular surface instanton rate
constants are illustrated in Fig.~\ref{fig:kie}.  The lines shown in that
figure are the fits discussed in detail in Sec.~\ref{sec:fit}.

\begin{deluxetable}{ccccccc}
\tablenum{3}
\tablecaption{The activation energies (ER mechanism), crossover temperatures and 
kinetic isotope effects~(KIEs) of the deuterium-substituted reactions 
corresponding to reactions~\ref{re:r1}, \ref{re:r2} and \ref{re:r3}. 
The energies and temperatures are given K.}\label{tab:tab3}
\tablewidth{0pt}
\tablehead{
\colhead{Reactions} & \colhead{$V_\text{act}$} & \colhead{$E_\text{act}$} &
 \colhead{$T_\text{c}$}&
\colhead{} & \colhead{KIEs} & \colhead{} \\
\cline{5-7}
\colhead{}&\colhead{}&%
\colhead{}& \colhead{} & \colhead{85~K} & \colhead{75~K} & \colhead{59~K}
}
\startdata
R(1) D+H$_{2}$CO & 1420 & 1750 &  141  &  22   &      &      \\
R(2) D+H$_{2}$CO & 5220 & 5470 &  280  & 3161  & 4192 &      \\
R(3) D+H$_{2}$CO & 4030 & 2700 &  365  &   7.9    &    9.6 & 12   \\
R(3) H+DHCO      & 4030 & 3540 &  288  &  136     & 145   & 165  \\
R(3) D+DHCO      & 4030 & 3210 &  273  &  631      & 880   & 1431 \\
\enddata
\end{deluxetable}

Compared with the reactions with a protium (H) atom, the ones involving
deuterium (D) always have much smaller tunneling rate constants. As frequently
observed for tunneling reactions, the KIEs for all three product channels
increased with decreasing temperature. The crossover temperature $T_\text{c}$
of the reactions with deuterium were reduced to 141~K for the
reaction~\ref{re:r1}, 280~K for the reaction~\ref{re:r2} and 365~K, 288~K,
 and 273~K for three cases of the reaction~\ref{re:r3} as shown in
Table~\ref{tab:tab3}. In that respect, H + DHCO refers to the reaction where D
is abstracted by an incomming H atom, D + DHCO means that D is abstracted by
an incomming D atom. In addition, this table provides activation energies and
KIEs at chosen temperatures as well. It reveals that the KIEs are strong for
all reactions and vary largely from tens to several thousands at low
temperatures.  Comparing the activation energies including ZPE
($E_\text{act}$) in Table~\ref{tab:tab2} and \ref{tab:tab3}, we can find the
differences of zero-point energies are not the main reason for strong KIEs
which implies tunneling may dominate this phenomenon.

\section{Implementation in astrochemical models\label{sec:fit}}

The rate constants calculated in this work can be directly used as input for
astrochemical models.  To facilitate that, we fitted the calculated tunneling
rates with the rate equation \citep{zhe10}:
\begin{equation}
  \label{equ:rate}
  k(T) = \alpha\left(\frac{T}{300\text{K}}\right)^{\beta} \exp\left(-\frac{\gamma(T+T_{0})}{T^{2}+T_{0}^{2}}\right)
\end{equation}
where $\alpha$, $\gamma$ and $T_{0}$ are fitting parameters while $\beta$ is a
constant set to one.  The pre-exponential factor $\alpha$ has the same unit as
the rate constant.  Both $\gamma$ and $T_{0}$ have the unit of
temperature. The parameter $\gamma$ is related to the height of the barrier,
while $T_{0}$ relates to the onset of strong tunneling.

\begin{deluxetable*}{cccccccccc}
\tablenum{4}
\tablecaption{Parameters for rate constants described of the reactions~\ref{re:r1}, 
\ref{re:r2} and \ref{re:r3} and the corresponding deuterium-substituted reactions 
by Eq.~(\ref{equ:rate}). 
 } \label{tab:tab4}
\tablewidth{0pt}
\tablehead{
\colhead{}    & \colhead{}    &  \multicolumn{3}{c}{\ce{H + H2CO}}     & 
 \multicolumn{3}{c}{\ce{D + H2CO}}          &
\colhead{\ce{H + DHCO}} & \colhead{\ce{D + DHCO}}       \\
\cline{6-8} \cline{10-10}
\colhead{}    & \colhead{}    & \colhead{R(1)}  & \colhead{R(2)} & \colhead{R(3)} & 
\colhead{R(1)}& \colhead{R(2)}& \colhead{R(3)}  &
\colhead{R(3)}& \colhead{R(3)}
}
\startdata
              & $\alpha$ ($10^{-12}$ cm$^3$s$^{-1}$) & 
$3.80$ & $2.26$ & $6.44$ & 
$2.54$ & $1.45$ & $6.11$ & 
$5.04$ & $3.14$ \\
      ER      &  $\beta$ & 1 & 1 & 1 & 1 & 1 & 1 & 1 & 1  \\
  mechanism   & $\gamma$ (K) & 1107  & 3635  & 1597  & 1112  & 3748  & 1538  & 2147  & 1918   \\
              & $T_{0}$ (K)  & 131.1 & 223.8 & 161.0 & 94.5  & 163.4 & 127.1 & 154.9 & 124.1  \\
 \hline
              & $\alpha$ ($10^{10}$ s$^{-1}$) & 
$3.14$ & $1.14$ & $4.13$ & 
$2.26$ & $0.75$  & $4.03$ & 
$3.44$ & $2.20$  \\
      LH      & $\beta$ & 1  & 1 & 1 & 1 & 1 & 1 & 1 & 1   \\
  mechanism   & $\gamma$ (K) & 830   & 3146  &  1222 & 903  & 3336   & 1226   & 1770  & 1605   \\
              & $T_{0}$ (K)  & 119.6 & 219.3 & 147.7 & 80.0 & 155.3  & 110.9  & 144.6 & 111.2  \\
\enddata
\end{deluxetable*}


Below the crossover temperature $T_{c}$, the instanton rate constants were
applied for fitting, while above $T_{c}$ until 1000~K, rate constants were
provided by modified transition state theory where vibrations were treated by
quantum harmonic oscillators and tunneling corrections were included by a
symmetric Eckart barrier with $\omega_\text{b}$ and $E_\text{act}$ taken from
the QM/MM calculations.  The fitted results were illustrated as lines in
Figs.~\ref{fig:instant} and \ref{fig:kie} and the resulting parameters are
listed in Table~\ref{tab:tab4}.  The lines are smooth and match the original
data well. We only provide fits to the tunneling rate constants for surface
reactions. The fits can be assumed to be reliable from high temperature to
approximately the lowest temperature, for which instanton rate constants were
available for the specific case, i.e. down to about 60~K for the reactions
with H transfer and down to about 40--50~K for reactions involving D-transfer.

For the ER process, the rate constants presented here can directly be used in
astrochemical models. In the LH process, there exists the competition between
reaction, diffusion out of the site, and desorption from the surface. 
The probability for reaction is expressed as
\begin{equation}
\label{equ:prob}
P_{\mathrm{react}}=\frac{k_{\mathrm{react}}}{k_{\mathrm{react}}+k_{\mathrm{diff}}+k_{\mathrm{desorp}}}
\end{equation}
where $k_{\mathrm{react}}$ is the $k(T)$ we provide in this work,
$k_{\mathrm{diff}}$ is the sum of the diffusion rate constants of the two
species, and $k_{\mathrm{desorp}}$ is the sum of the desorption rate constants
of the two species. All three of these quantities are unimolecular rate
constants. The reaction rate of the LH process, used in astrochemical models
is 
\begin{equation}
\label{equ:lh}
R_{\mathrm{LH,react}}=P_{\mathrm{react}}R_{\mathrm{diffusion}}.
\end{equation}
Here, $R_{\mathrm{diffusion}}$ is the rate by which two species meet on the
surface due to diffusion:
\begin{equation}
\label{equ:diff}
R_{\mathrm{diffusion}} = \frac{n_{\mathrm{A}}n_{\mathrm{B}}}{n_{\mathrm{sites}}}k_{\mathrm{diff}}
\end{equation}
where $n_\text{X}$ is the concentration of species X adsorbed on the surface
and $n_{\mathrm{sites}}$ is the concentration of adsorption sites. These
expressions are valid for $n$ being a volume concentration, a surface
concentration, or a number density. 

Our data provide $k_\text{react}$. If the reaction on the surface in the LH
mechanism is faster than diffusion of H atoms to the reactant, the reaction is
diffusion-limited. In that case, $R_\text{LH,react}=R_\text{diffusion}$ is
independent of $k_\text{react}$. Then the absolute rate constant will be
different from those reported here, but the branching ratio between the three
channels should be largely unaffected, \ref{re:r1} is still expected to
dominate the surface process. The binding modes, and consequently blocking of
routes for attack, might influence the branching ratio. Here, we only studied
one binding site in detail. Since the vast majority was found in binding type
(a), which allows access to all three routes, that surface influence can be
expected to be small.


\section{Conclusion \label{sec:concl}}

In this paper, we studied three product channels of the reaction H +
H$_{2}$CO, producing CH$_{3}$O, CH$_{2}$OH, and H$_{2}$ + HCO on the ASW
surface. QM/MM modeling was used combined with instanton tunneling
calculations.  Three types of binding modes of H$_{2}$CO on the ASW surface
were found. The binding energies had a broad distribution from $\sim$ 1000~K
to 9370~K.  The activation barriers were found to be independent of the
binding energies.  We presented both unimolecular and bimolecular instanton
tunneling rate constants for those three product channels. In all three
reactions, we found a significant contribution of quantum tunneling at low
temperature. It turned out that the ASW surface has a noticeable catalytic
effect, reversing the branching ratio between reactions \ref{re:r1} and
\ref{re:r3}. The ASW surface enhances the formation of CH$_{3}$O and hinders
hydrogen abstraction to form H$_{2}$ + HCO. The channel leading to CH$_{2}$OH
is negligible in all cases.  In addition, strong kinetic isotope effects were
found in all three channels differing between tens to several thousands at low
$T$.  Finally, we fitted our calculated rate constants to a modeler-friendly
form and provide the fitted parameters for future astrochemical applications.
 
\acknowledgments
This work was financially supported by the European Research Council (ERC)
under the European Union’s Horizon 2020 research and innovation programme
(grant agreement No 646717, TUNNELCHEM). The authors also acknowledge support
for CPU time by the state of Baden-W\"urttemberg through bwHPC and the German
Research Foundation (DFG) through grant no INST 40/467-1 FUGG.

\bibliography{h2co_ver1}

\appendix
\section{Data for the Benchmark Calculations\label{sec:databen}}

Results of the benchmark calculations are given in Table~\ref{tab:tab1}. The
functional/basis set combination PWB6K-D3/def2-TZVP resulted in activation
energies closest to the (U)CCSD(T)-F12/cc-pVTZ-F12 reference data and was,
therefore, chosen for the present study.

\begin{deluxetable}{llrrr}
\tablenum{1}
\tablecaption{Benchmark calculations for the reactions~\ref{re:r1}, \ref{re:r2} and   
\ref{re:r3}.}\label{tab:tab1}
\tablewidth{0pt}
\tablehead{
\colhead{Functional} & \colhead{Basis-set} &  
\multicolumn{3}{c}{$V_\text{act}$ (K)}  \\
\cline{3-5}
\colhead{} & \colhead{} & \colhead{R(1)} &
\colhead{R(2)} & \colhead{R(3)}
}
\startdata
(U)CCSD(T)-F12 & cc-pVTZ-F12    & 1570 & 4580 & 3300 \\
PWB6K-D3       & def2-TZVP  & 1660 & 4670 & 3200 \\
PWB6K-D3       & def2-SVPD  & 1400 & 4280 & 3210 \\
B1B95-D3       & def2-TZVP  &  1150 & 3500 & 1510 \\
B1B95-D3       & def2-SVPD  &  830 & 3040 & 1500 \\
B3LYP-D3       & def2-TZVP  &  40 & 1660 & $-$330 \\
B3LYP-D3       & def2-SVPD  & $-$300 &  1170 & $-$290 \\
BECKE97-D-D3   & def2-TZVP  &  780 & 1850 &$-$1670 \\
BECKE97-D-D3   & def2-SVPD  &  500 & 1490 &$-$1510 \\
BHLYP-D3       & def2-TZVP  &  820 & 3650 & 2320 \\
BHLYP-D3       & def2-SVPD  &  590 & 3310 & 2430 \\
MPWB1K-D3      & def2-TZVP  & 1480 & 4400 & 2810 \\
MPWB1K-D3      & def2-SVPD  & 1220 & 4000 & 2830 \\
MPW1B95-D3     & def2-TZVP  & 1160 & 3560 & 1740 \\
MPW1B95-D3     & def2-SVPD  &  840 & 3100 & 1730 \\
PBE0-D3        & def2-TZVP  &  550 & 2420 &  1030 \\
PBE0-D3        & def2-SVPD  &  310 & 2010 &  1070\\ 
PBE96-D3       & def2-TZVP  & $-$270 &  620 & $-$1190 \\
PBE96-D3       & def2-SVPD  & $-$670 &   30 & $-$1310 \\
PW6B95-D3      & def2-TZVP  &  960 & 3310 & 1390 \\
PW6B95-D3      & def2-SVPD  &  620 & 2830 & 1360 \\
SSB-D-D3       & def2-TZVP  &   40  & 1480 & $-$570 \\
SSB-D-D3       & def2-SVPD  & $-$380 &  1020 & $-$680 \\
TPSS-D3        & def2-TZVP  &$-$2070 & $-$600 &$-$2530 \\
TPSS-D3        & def2-SVPD  &$-$2280 & $-$1020 &$-$2490 \\
TPSSH-D3       & def2-TZVP  &$-$1760 &  80 &$-$1730 \\
TPSSH-D3       & def2-SVPD  &$-$1930 & $-$310 &$-$1670 \\
\hline
(U)CCSD(T)-F12\tablenotemark{a} & cc-pVTZ-F12    & 1770 & 4910 & 3540 \\
PWB6K-D3\tablenotemark{a}       & def2-TZVP  & 1620 & 4780 & 3320 \\
\enddata
\tablenotetext{a}{Structures from the geometry optimization based on 
PWB6K-D3/def2-TZVP theory level}
\end{deluxetable}

\end{document}